\documentstyle[pre,aps,multicol,epsfig,color]{revtex}
  
\begin{document}
\draft


\title{Analysis of a spatial Lotka-Volterra model with a finite  range predator-prey interaction}

\author{ E. Brigatti$^{\dag}$, M. N\'u\~{n}ez-L\'opez$^{\pm}$ and M. Oliva$^{\ddag}$}

\address{
$\dag$ Instituto de Ciências Exatas, Universidade Federal Fluminense - Rua E. H. Figueira, 783, 
RJ, Brasil \\
  Instituto de F\'{\i}sica, Universidade Federal Fluminense - Campus da Praia Vermelha, 24210-340, Niter\'oi, RJ, Brasil\\
$\pm$ Instituto Mexicano del Petr\'oleo -
Eje Central L\'azaro C\'ardenas Norte 152, Gustavo A. Madero, 07730
DF, M\'exico\\  
$\ddag$ Facultad de F\'{\i}sica, Universidad de La Habana - Ave. Universidad y Ronda, Vedado, 10400, Havana, Cuba 
}


\maketitle 
\widetext

\begin{abstract}
We perform an analysis of a recent spatial
version of the classical Lotka-Volterra model, where a finite
scale controls individuals' interaction. We study the
behavior of the predator-prey dynamics in physical spaces higher
than one, showing how spatial patterns can emerge for some values
of the interaction range and of the diffusion parameter. 
\end{abstract}


\begin{multicols}{2}



\section{Introduction}

Reaction-diffusion theory plays a very important role in the
study of pattern formation in biology, an aspect of populations
dynamics  which has gained importance since some Moran's seminal
works \cite{moran,lieb}. More recently,  theoretical ecology studies
focused on the dynamical mechanisms which can be considered
responsible for producing such structures. Their principal
aim is to determine the local rules which cause the emergence
of macroscopic patterns \cite{biospace}. At present,
a central issue 
is to determine whether the predator-prey interactions can be
considered among these mechanisms. Actually, spatial correlations
between predators and preys populations have been observed in real
systems such as predatory beetles and larval flies as their
preys\cite{tobin1}. This research shows that these larval flies
appear to gain local clustering as beetle abundance approaches a
carrying capacity with the prey population.

A traditional approach for describing these systems
is based
on spatial Lotka-Volterra type models. This
description is able to generate spatial structures of the Turing type, where
the instabilities are directly driven by mechanisms connected with the
process of diffusion \cite{murray89}. 
These works are inspired to
the classical ideas introduced by Turing, who showed that diffusion can destabilize
the homogeneous equilibrium solution of a reaction-diffusion system \cite{Turing}.
This topic has recently received some attention,
in particular in relation to the effects of stochasticity.
For example, in \cite{physrev2} it is shown that Turing-like patterns can indeed emerge beyond the parameter region predicted by the conventional Turing theory.
This phenomenon is due to the effect of finite size corrections on the mean-field idealized dynamics.
Also demographic noise can induce persistent spatial pattern formation and temporal oscillations \cite{physrev1}. Specifically, demographic noise greatly enlarges the region of parameters space where pattern formation occurs, in relation to the 
prediction of the mean-field theory.
Finally, Scott et. al. \cite{Scott} introduce an analytic method to describe
stochastic spatially varying systems governed by reaction-transport master equation.

More recently, a different mechanism capable of generating spatial patterns has been explored \cite{epl}. It is based on the introduction of
a probability of interaction (predation) which
becomes a function of the distance between individuals.
This description presents two interesting points.
First, it is not subject to the strong constraints
which limit the emergence of spatial inhomogeneities in
the traditional approach \cite{holmes}.
Second, it is based on
a simple ingredient which has a deep
ecological motivation.
In fact, it has became evident that predation
strongly depend on
probability of encounter \cite{tobin2} which must take into account
predators shifts in response to prey movements. 
 Moreover, territoriality is another fundamental aspect in the ecology of many 
predatory animals such as wolves, lions, hyenas, African wild dogs and badgers. 
In this case an immediate question arises as to how the predation is strongly influenced
by a land which is divided up into predator territories \cite{murray89}.
These elements
result in a spatial and temporal variation in the predation risk \cite{Lima}.
A possible description of one aspect of this spatial dependence
can be obtained
considering that the probability at which a consumer meets a prey should be dependent
on the relative distance between them \cite{ecology}. The
introduction of a spatial scale of interaction has been widely
applied to model competition in
community ecology  \cite{ecology} and more recently it was also used
in studies of  evolutionary theory \cite{speciation}.

In the following, we present a detailed analysis of this
approach for an implementation of a predator-prey system in a two
dimensional physical space, the most relevant for describing real ecosystems.

\section{The mean-field model}

We start with the mean-field version of the model.
This description is characterized by a couple of equations, one for the
preys $N(\bar{x},t)$  and other for the predators $P(\bar{x},t)$. They
describe diffusion in real space and the strength of the
interaction in the nonlinear term is a function of the density of individuals
in the proximity \cite{lopez}. 
The reason for introducing a spatial scale for the interaction 
can be clearer understood considering that, based on the reported
ecological motivations, we want to describe a probability of
reproduction/death  not constant but dependent  on the number of predators/preys 
in the surrounding. 
In fact, in a region more populated by preys, predators should have a better chance of reproduction. 
Conversely, in a region more populated by 
predators, preys should have a greater chance of dying. 
This nonlocality appears as an integral term
which takes into account the local density of population.
It is important to point out that this effect is not related to
the individuals' velocity, controlled by the 
diffusion coefficients, but to the memory which individuals maintain 
in relation to a specific territory. 
Territoriality is a fundamental aspect in the ecology of many 
predatory animals \cite{murray89}. 
For this reason, as the range of interaction is not simply linked to 
the relative motion between predators and preys, it does not depend on a single interaction 
scale.

Considering a two dimensional physical space the equations which describe our system
are the following:
{\scriptsize
\begin{eqnarray}
\frac{\partial N(\bar{x},t)}{\partial
t}&=&D_{N}\frac{\partial^{2}N(\bar{x},t)}{\partial \bar{x}^{2}} +rN(\bar{x},t) \nonumber \\
  & &-\alpha N(\bar{x},t) \int_{|\bar{x}'-\bar{x}|<R_{1}} P(\bar{x}',t) d\bar{x}' \nonumber \\
\frac{\partial P(\bar{x},t)}{\partial
t}&=&D_{P}\frac{\partial^{2}P(\bar{x},t)}{\partial \bar{x}^{2}} -mP(\bar{x},t) \nonumber \\
  & &+\beta P(\bar{x},t) \int_{|\bar{x}'-\bar{x}|<R_{2}} N(\bar{x}',t) d\bar{x}'.\nonumber \\
\label{eq_1}
\end{eqnarray} }
\noindent where $\bar{x} = (x,y)$ is the position of predators or preys.
Predators consume the preys with an intrinsic rate $\alpha$ and
reproduce with rate $\beta$, 
$r$ is the prey's growth rate and predators are assumed to spontaneously
die with rate $m$. 
$D_N$ and $D_P$ are the diffusion coefficients of preys and predators, respectively.

A microscopic version of these equations is introduced in the next section.
There, the integral terms correspond to counting the number of preys/predators 
which are at a shorter distance than $R_i$ from the considered predator/prey.

We would like to point out that a rigorous derivation, which would show that the continuum 
field equations introduced here are totally analogous to the discrete model presented in the next section,
can be obtained by using Fock space techniques after making some usual assumption and approximation \cite{lopez,tauber}.

To perform this passage between discrete model and mean field description 
the microscopic model should be defined on a lattice.
The lattice model will be equivalent to the off-lattice one in the limit of small lattice size.
For this reason the off-lattice model present in the next section should be modified to fit into a grid. 
If we let the grid spacing to become negligibly small we recover the continuum limit. 
With the aim of preserving  the non-locality of our macroscopic model, 
in which the interaction range is finite, this operation of limit should be made carefully.
In fact, only the grid size, but not the interaction lengths $R_i$ should vanish. 
This is obtained fixing $R_i$ to a finite value when going to the continuum.

Analogous models, but not characterized by a finite scale of interaction, appeared recently in the literature.
Spatio-temporal patterns can emerge sustained by the stochastic component of the dynamics.
In this case, the demographic fluctuations, directly reproduced in the individual based description 
level, are the responsible for pattern formation \cite{physrev1,Anna}.

The one dimensional case was deeply analyzed in the reference \cite{epl}.
In the following we will present some results related to the two-dimensional
physical space, and, for theoretical interest, to n-dimensional spaces.

Also in the two dimensional space, an absorbing phase
$N(\bar{x},t)=P(\bar{x},t)=0$ and a survival phase
$\bar{N}(\bar{x},t)=\frac{m}{\pi\beta R^2_{2}}$;
$\bar{P}(\bar{x},t)=\frac{r}{\pi\alpha R^2_{1}}$ exist and they
correspond  to spatially homogeneous, stationary
solutions.

The analysis for detecting the presence of solutions with
spatial structure is based on the idea of perturbing
the survival phase stationary solution $\bar{N}(\bar{x},t)$ and
$\bar{P}(\bar{x},t)$. This is obtained introducing small harmonic perturbations
\cite{murray89,lopez}: $N(\bar{x},t)=\bar{N}+A_{N}exp[\lambda t +i
\bar{k}\cdot \bar{x}]$; $P(\bar{x},t)=\bar{P}+A_{P}exp[\lambda t+i \bar{k}\cdot
\bar{x}]$, where $\bar {k} = (k_x , k_y)$
is a two-dimensional wave vector of modulus $|{\bar{k}}|=k$.
The simple harmonic form of these perturbations
reduces equations \ref{eq_1} to a linear system.
The solution of this system generates
some constraints on the values which
$\lambda$ and $k$ can assume.
This dispersion relation can be written in the following form:
{\scriptsize
\begin{eqnarray}
\lambda(k) &=& -\frac{k^{2}}{2}(D_{N}+D_{P}) \nonumber \\
           & & \pm \sqrt{\frac{k^{4}}{4}(D_{N}-D_{P})^{2}-4rm
\frac{J_1(kR_{1})J_1(kR_{2})}{k^{2}R_{1}R_{2}}}, \label{eq:2}
\end{eqnarray}}
\noindent where $J_1$ is the first-order Bessel function.

If, for some $k$, $Re[\lambda(k)] > 0$ spatial patterns can emerge.
It follows that $R_{1}\neq R_{2}$ is a necessary condition.
Moreover, because for  $D_{N}=D_{P}=0$ the condition
is always satisfied, it is evident that the instability is driven by the
range of the interaction and is independent of the diffusion
process.

\begin{figure}
\centerline{\psfig{figure=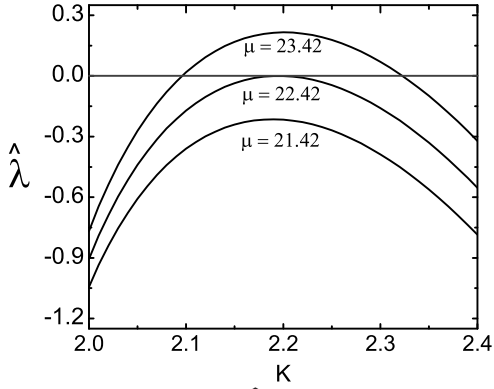, width=6.5cm, angle=0}} \caption{
\small  Dispersion relation $\widehat{\lambda}(K)$ for different
values of the parameters ($\mu=\sqrt{rm}R^{2}/D$).}
\label{fig_dis}
\end{figure}

If we consider the simpler situation with
$D_{N}=D_{P}=D$ and $R_{1}=2R_{2}=2R$, and
we introduce the rescaled variables $K=kR$ and
$\widehat{\lambda}=\lambda\frac{R^{2}}{D}$, for
$\mu=\sqrt{rm}R^{2}/D$ equation~\ref{eq:2} reduces to: {\scriptsize
\begin{equation}
\widehat{\lambda}(K)=-K^{2}+\frac{\mu}{K}\sqrt{-J_{1}(K)J_{1}(2K)}.
\label{eq:3}
\end{equation}}
The edge of spatial patterns emergence
corresponds to the values of the parameters for which
the maximum  $K_{m}$ of $\widehat{\lambda}(K)$ becomes
positive.
For the original variables this happens if:

\begin{equation}
\qquad \frac{\sqrt{rm}R^{2}}{D}\gtrsim 22.4228 \qquad \textrm{and } \qquad k_{m}\approx \frac{2.19535}{R}.
\label{eq:4}
\end{equation}
Figure \ref{fig_dis} shows $\widehat{\lambda}(K)$ for different $\mu$ values.

In the following we turn our attention to the model implemented in a $n$-dimensional space,
with $n>2$.
In this case, the corresponding equations are obviously obtained generalizing the
Laplacian terms and
the integration domains.
Performing a calculation analogous to the 2-dimensional case,
we arrive at the following dispersion relation:
{\scriptsize
\begin{eqnarray}
\lambda(k) &=& -\frac{k^{2}}{2}(D_{N}+D_{P}) \nonumber \\
           &  \pm&
           \sqrt{\frac{k^{4}}{4}(D_{N}-D_{P})^{2}-rm 2^{n}\Gamma \left( 1+ \frac{n}{2}\right)^{2}\frac{J_{n/2}(kR_{1})J_{n/2}(kR_{2})}{(kR_{1})^{n/2}(kR_{2})^{n/2}} },
          \nonumber \\
\end{eqnarray}}

\noindent where $n$ is the space dimension.
Introducing the variables $K=kR_{2}$,
$\rho=\frac{R_{1}}{R_{2}}$ and
$\mu=R_{1}R_{2}\sqrt{\frac{rm}{D_{N}D_{P}}}$,
spatial patterns can emerge if the following condition is satisfied:

\begin{equation}
\frac{\rho^{2}K^{4}}{\mu^{2}}+2^{n}\Gamma\left(1+\frac{n}{2}\right)^2
\frac{J_{n/2}(K)J_{n/2}(\rho K)}{\rho^{n/2}K^{n}}<0.
\label{eq_rel}
\end{equation}

\begin{figure}
\centerline{\psfig{figure=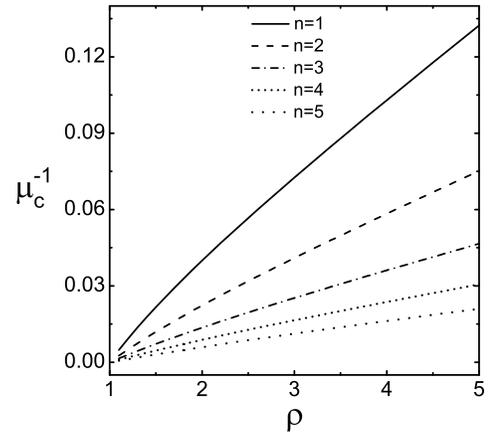, width=6.3cm, angle=0}}
\caption{ \small  $\mu_c^{-1}$ values in dependence of $\rho$
for different physical dimensions $n$.} \label{fig_phy}
\end{figure}

\noindent
As for the two dimensional situation, a  critical
value $\mu_{c}$ exists for which, if $\mu>\mu_{c}$ spatial
patterns emerge.
In Figure~\ref{fig_phy} we show the $\mu_{c}^{-1}$
values as a function of the ratio $\rho=\frac{R_{1}}{R_{2}}$ for different dimensions.
The value of $\mu_{c}^{-1}$ increases as the parameter $\rho$ increases.
This means that a stronger difference between $R_1$ and $R_2$ leads to easier cluster formation.
Moreover, it is clear that for $\rho>1$
clustered distributions can appear also for a physical dimension greater than $2$.
In higher dimensions cluster formation becomes more difficult but
it is viable.
It is important to remember that even the case of a 3-dimensional space
is not a common ecosystem. Obviously it does not exist for terrestrial
individuals and even marine species tend to remain within a layer of small
thickness compared to their horizontal movements. Anyway, our
results for dimension higher than $2$ have a theoretical interest. In fact,
qualitative differences depending on the space dimension are  a common
feature in systems with diffusive phenomena and clusters formation (see for example \cite{Houch}).

\section{The microscopic discrete model}

In the following we introduce a microscopic discrete stochastic
formulation of the mean field model.
The importance of comparing the results of the mean-field
description with an individual based one is due to
the possibility of introducing demographic fluctuations.
These fluctuations result in the appearance
of an intrinsic stochasticity which,  in
principle, is always present in real phenomena.
Moreover, this technique
can describe the occurrence of threshold effects,  generated by the discrete
nature of individuals. These effects are not present in the
mean field description where every small amount of the density of
population is acceptable, even if unrealistically small  \cite{attofox}.

We implement our individual based model on the square
$[0,1]\times[0,1]$, with periodic boundary conditions. Simulations
start with an initial population of $P_{0}$ predators and  $N_{0}$
preys, randomly located.
A time step of our simulation corresponds to carrying
out the following processes:
1) diffusion, where a predator, or a prey, is randomly selected and
moves some distance, in a random direction, chosen from a Gaussian
distribution of standard deviation $\sigma$. 
2) predator reproduction, with rate $\beta
N^{\bar{x}}_{R_{2}}$, where $N^{\bar{x}}_{R_{2}}$ is the number of preys which
are at a shorter distance than $R_{2}$ from the predator at
position $\bar{x}$. 3) predator death, with probability $m$. 4) prey
reproduction, with probability $r$. 5) prey death, with rate
$\alpha P^{\bar{y}}_{R_{1}}$, where $P^{\bar{y}}_{R_{1}}$ is the number of
predators which are at a shorter distance than $R_{1}$ from the
prey at position $\bar{y}$. All the newborns maintain the same location
as the parents. We evaluate $N^{\bar{x}}_{R_{2}}$ and $P^{\bar{y}}_{R_{1}}$
using periodic boundary conditions.
These processes are executed sequentially by the whole population.
In fact, for each action, an
individual of each population (predator or prey) is randomly selected
and this operation is repeated for a number
of times equal to the size of the corresponding population.

It is important to point out that if we
measure time in units of the simulation time step, the coefficient
$D$ of equation \ref{eq_1} is related to the discrete model through the
relation $D=\sigma^{2}/2$.
Birth and death probabilities  are the same in the continuous and
in the discrete model.

\begin{figure}
\centerline{\psfig{figure=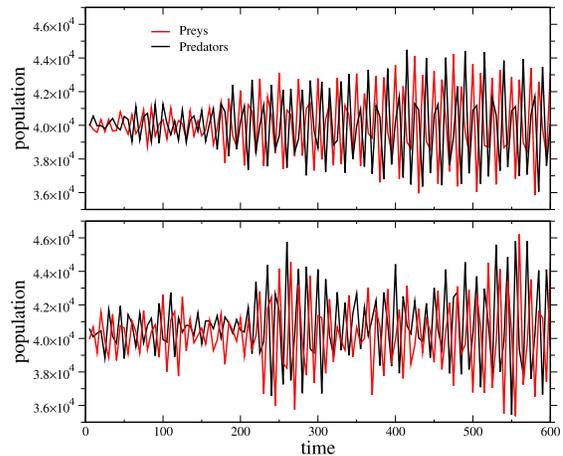, width=8.cm, angle=0}}
\caption{\small Temporal evolution for populations with homogeneous spatial distributions.  Top: $R_{1}=2 R_{2}= 0.16$, 
$\sigma=0.03$, $r=m=0.5$, $\alpha=6.22\times 10^{-4}, \beta=1.55\times 10^{-4}$, $P_{0}=N_{0}=40000$.
Bottom: $R_{1}=R_{2}= 0.08$, $\sigma=0.008$, $r=m=0.5$, $\alpha=\beta=6.22\times 10^{-4}$, $P_{0}=N_{0}=40000$. The two temporal evolutions are very similar. This analogous behaviour is caused by the fact 
that no patterns are generated in either of the two simulations: in the upper one because of the large value of $\sigma$, in the lower one because of $R1 = R2$.
} \label{fig_tempo1}
\end{figure}

Analyzing the data generated throughout these simulations
we are able to explore the temporal and spacial behavior of our
system.
The temporal dynamics of our simulations is strongly
dependent on the form of the spatial organization of the
populations. The temporal evolution of the simulations
characterized by homogeneous spatial distributions can be easily
understood considering a classical Lotka-Volterra model.
In fact, if we start with an initial population of
$P_{0}=\frac{r}{\pi\alpha R^{2}_{1}}$ predators and
$N_{0}=\frac{m}{\pi \beta R^{2}_{2}}$ preys, corresponding to the two
stationary solutions of the mean field model,  inevitable
fluctuations tend to push the system away from the trivial
survival phase and induce irregular population oscillations that
almost resemble the deterministic cycles of the classical
Lotka-Volterra model. This behaviour has been described also in
other analogous  stochastic models \cite{lattice,tauber}.
We must
remember that an important outcome of the introduction of
intrinsic stochasticity is the possibility of predators' extinction.
This fact, combined with the possibility of extremely rapid increase
in the number of preys, force to a careful choice of
the initial conditions and parameters of our simulational runs.
This is necessary to obtain controlled irregular oscillations, which swing in a rather erratic
fashion around an average value (see Figure~\ref{fig_tempo1}).
Otherwise, the explosion or extinction of population do not
allow to carry out sufficiently long standing simulations.

\begin{figure}
\vspace*{0.8cm}
\centerline{\psfig{figure=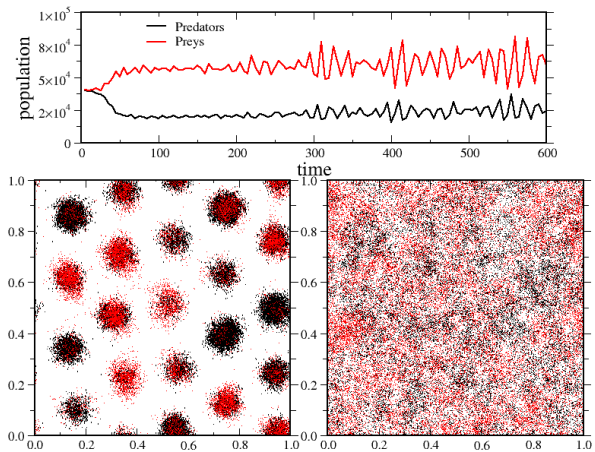, width=8.cm, angle=0}}
\vspace*{0.4cm}
\caption{\small Top: temporal evolution for populations with modulated
spatial distributions ($R_{1}= 2R_{2}= 0.16$). Other
parameters are: $\sigma=0.008$, $r=m=0.5$, $\alpha=6.22\times 10^{-4}, \beta=1.55\times 10^{-4}$,
$P_{0}=N_{0}=40000$. Bottom: on the left, modulated spatial distribution corresponding to the
simulation shown on the top of the figure. On the right, homogeneous distribution generated using the following parameters:  $R_{1}=R_{2}= 0.08$, $\sigma=0.008$, $r=m=0.5$, $\alpha=\beta=6.22\times 10^{-4}$, $P_{0}=N_{0}=40000$. }
\label{fig_mod}
\end{figure}

The situation changes when the system is evolving in the spatial
clustered regime. In fact, for some values of the parameters,
the survival phase of the homogeneous distribution is different from
the clustered one. For example, starting
with  $P_{0}=\frac{r}{\pi\alpha R^2_{1}}$ predators and
$N_{0}=\frac{m}{\pi\beta R^2_{2}}$ preys,
after a rapid transient, when spatial structures emerge, the
oscillations organise around a new survival phase. In fact, the
difference between $P^{\bar{y}}_{R_{1}}$ and $N^{\bar{x}}_{R_{2}}$ becomes
irrelevant and the values of $\alpha$ and $\beta$ determine the
ratio in the new population equilibrium (see
Figure~\ref{fig_mod}). This behaviour can be easily understood considering that the
dimension of the  bulk of the clusters
is generally smaller than $R_i$, as shown at the end of this section.
We can also remark that the more the  spatial solution is
marked by clustering (lower $R$ and $D$ values), the more the
irregularities of the temporal oscillations increase.
It is interesting to point out a recent model which shows analogous
time oscillations in predators and preys number associated with
a spatial structure.
These oscillations are due to stochastic fluctuations about
the time-independent solutions of the deterministic equations \cite{Lugo}.
Other models describe similar oscillating behaviour \cite{Dauxois}.
Interestingly, these works identify and analytically describe a mechanism of 
amplification of demographic noise which can give rise to coherent oscillations. 
In particular, as in our model, in references \cite{physrev1,Anna} these oscillations 
are spatiotemporal in nature.

In the following we present some results related to the
spatial distributions generated by our model.
As reported in Figure~\ref{fig_mod} regular patterns can be
obtained only for simulations where the two ranges
 of interaction, $R_{1}$ and $R_{2}$, have a different value.
This outcome is in agreement with the results obtained by
the mean field description.
The spatial patterns are characterised by a sequence of isolated colonies,
generally called of spikes \cite{lopez,edo}.
These spikes keep changing along the
time, according to the oscillation reported in the temporal
evolution of the total population (top of Figure~\ref{fig_mod}).
These dynamics result in periodic stationary spatial structures
characterised by fluctuating clusters arranged on an hexagonal lattice.
We can remark that changing the geometry of the domains of integration
it is possible to generate different clusters
arrangements (for example, a square domain generates
clusters arranged on a square lattice).

\begin{figure}
\centerline{\psfig{figure=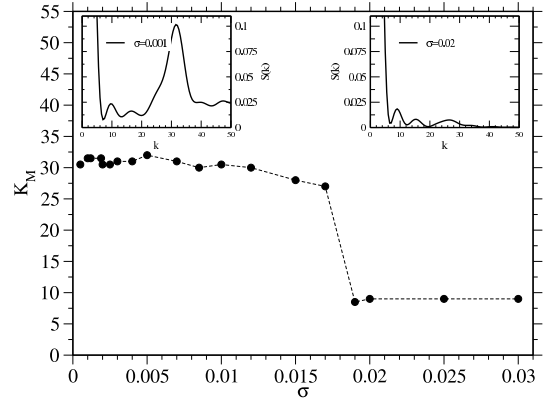, width=8.cm, angle=0}}
\caption{\small $K_{M}$ as a function of $\sigma$ ($R_{1}=2R_2= 0.16$, 
$r=m=0.5$, $\alpha=6.22\times 10^{-4}, \beta=1.55\times 10^{-4}$,
$P_{0}=N_{0}=40000$).
For this parameters the analytical
prediction gives $\sigma_{c}=0.019$, in good accordance with
the data obtained from the discrete model.}
\label{fig_trans1}
\end{figure}

In the following, we characterise the transition between the
homogeneous spatial distribution and the inhomogeneous one
(segregation transition). A proper order parameter is obtained
evaluating a structure factor $S(K)$ defined as:
\begin{equation}
S(K)=\Big\langle\Big| \sum
\limits_{j=1}^{N(\tau)}\frac{1}{N(\tau)}\exp[i  {{\bar q}\cdot
{\bar x}}_{j}(\tau)]\Big|^2\Big{\rangle}_{K}
\end{equation}
\noindent where the sum is carried
out over all predators (or preys) $N(\tau)$ at time $\tau$,
 ${\bar x}_j(\tau)=(x_j(\tau),y_j(\tau))$ is the position of the predator $j$, ${\bar q} = (q_x , q_y)$
 is a two-dimensional wave vector and the average is a spherical average over all wave vectors of modulus $|{\bar q}|=K$.
The position of the secondary local maximum of this function ($K_{M}$)
identifies the emergence of periodic structures in the spatial distribution.
In fact, the transition from a homogeneous to an inhomogeneous
distribution matches the jump of $K_{M}$ to higher values,
corresponding to the wave number of the periodic clusters present
in the space. This is clearly shown in the insets of Figure~\ref{fig_trans1}.
The segregation transition is characterised by the passage of $K_{M}$ from
small values to higher values as soon as a modulation becomes dominant
\cite{lopez,edo}.

In Figure~\ref{fig_trans1} we show $K_{M}$ as a function of $\sigma$.
As predicted by the mean field approximation, for
any value of the range of the interaction, a critical value of the
diffusion coefficient 
exists above which no spatial
structures emerge. In the figure we show some results obtained
fixing the parameters $R_{1}=2R_{2}=0.16$.
For the sake of simplicity we explored the case with  $R_{1}=2R_{2}$, but similar behaviour 
can be obtained for other values of $\rho\ne1$.

\begin{figure}
\centerline{\psfig{figure=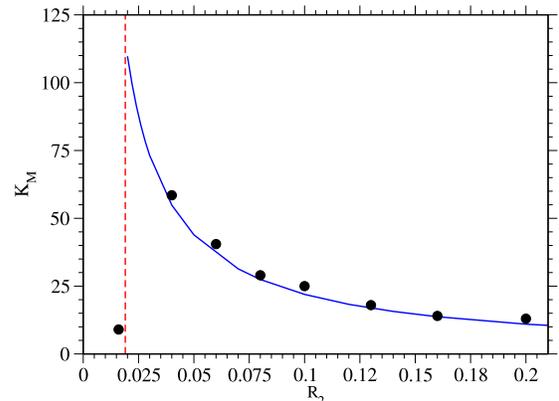, width=8.cm, angle=0}}
\caption{\small $K_{M}$ as a function of $R_{2}$
($\sigma= 0.004$, $r=m=0.5$,$R_{1}=2 R_{2}$,
$P_{0}=N_{0}=80000$, $\alpha=[2\pi P_{0} R^{2}_{2}]^{-1}$, $\beta=0.25\alpha$).
Simulational data are compared with the analytical predictions given by eq.~\ref{eq:4},
from which we obtained:
$k_{m}\approx2.2/R_{2}$ (continuous line) and $R_{c}\approx0.019$ (dashed line). } \label{fig_trans2}
\end{figure}

A similar analysis can be unfolded fixing $\sigma$ and looking for the
$R$ dependence.
We obtained spatial modulations of arbitrary wavelengths
just tuning the parameter $R_{2}$.
In Figure~\ref{fig_trans2} we
can see how the continuous description can reproduce
quantitatively the period of the patterns which emerges from the
modulated distributions of the Monte Carlo simulations.
In fact, the wavenumber of these distributions are well
approximated by the second relation in equation~\ref{eq:4}, which gives
the fastest growing mode of the mean-field approximation.
For extremely short-ranged interaction, a noisy spatially homogeneous
distribution appears.
This means that  clusters appear  for $R_{2}>R_{c}$, where $R_{c}$ is a
critical value of the interaction length for which the segregation
transition takes place.

It is interesting to remark another phenomenon.
Also in a two dimension implementation, independently of the values of $R$,
for low $D$ values and low population
density the demographic fluctuations can introduce a source
of spatial correlation which can generate disordered clusters \cite{epl}.\\

\begin{figure}
\centerline{\psfig{figure=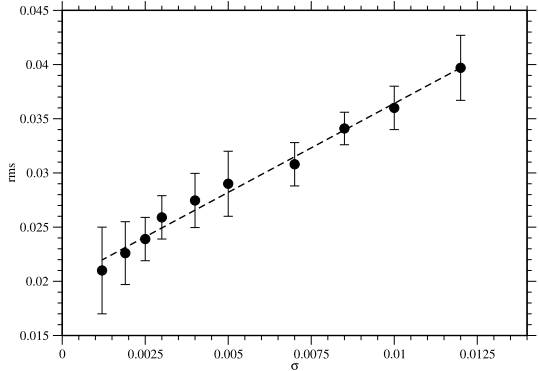, width=8.cm, angle=0}}
\caption{\small Cluster size as a function of $\sigma$. The dashed
line has slope 1, which corresponds to a square root dependence
of the cluster size on the diffusion coefficient.}  
\label{fig_clus}
\end{figure}

Finally, we try to evaluate  the
typical size $S$ of the clusters which appear in the spiky phase.
The typical cluster size was calculated averaging the root mean square
dispersion for all the clusters present in the system at a given
time. Data exposed in Figure~\ref{fig_clus} show a dependence
of the cluster size on the diffusion coefficient: $S \propto
\sigma$, equivalent to $S \propto \sqrt{D}$. 
We can easily interpret this result.
Since there is no attraction or repulsion between individuals, they
will experience just a diffusive motion during their lifetimes.
Assuming that the individuals confined in a cluster
diffuse a distance proportional to  $\sqrt{D}$, the 
dependence of the clusters size on the diffusion coefficient 
shown in Figure~\ref{fig_clus} is straightforward.
Similar results were reported for simpler models in \cite{edo,lopez3}.

 \section{Conclusions}

We have developed an in-deep analysis of a spatial Lotka-Volterra model
characterised by a finite range predator-prey interaction in
a two dimensional space. This physical space is obviously
the most appropriate for describing real ecosystems.

This model can be considered relevant from an
ecological perspective  for two reasons.
First, because it contributes
to the discussion about the mechanisms that can generate
spatial correlations in population ecology, showing how preys and
predators interaction can be considered among them. This
conjecture is of particular importance as some records of spatial
correlations between preys and predators were clearly
reported \cite{lieb,tobin1}.
Second, because the
emergence of spatial patterns in an erratic oscillatory regime,
which can contemplate predators extinction,  shows realistic
elements generally absent from conventional descriptions.

A theoretical interest related to this kind of
studies exists too. From this perspective,
we demonstrate that in this model instability is driven by
the form of the interaction and is independent of the diffusion
process.
Moreover, we carried out our analysis
developing a direct comparison between
an individual based implementation
and a mean-field
description, showing a perfect match between them
for many quantitative features,
a fact that is not always achieved \cite{physrev1,Lugo,edo,prl}.
Finally, we show how the presence of structures persists independently of
the space dimension where the model can be implemented.

\end{multicols}

\end{document}